\documentclass[twocolumn]{aastex631}

\accepted{August 30, 2024}

\begin{document}

\title{JADES: Spectroscopic Confirmation and Proper Motion for a T-Dwarf at 2 Kiloparsecs}

\author[0000-0003-4565-8239] {Kevin N.\ Hainline}
\affiliation{Steward Observatory, University of Arizona, 933 N. Cherry Ave, Tucson, AZ 85721, USA}

\author[0000-0003-2388-8172] {Francesco D'Eugenio}
\affiliation{Kavli Institute for Cosmology, University of Cambridge, Madingley Road, Cambridge CB3 0HA, UK}
\affiliation{Cavendish Laboratory, University of Cambridge, 19 JJ Thomson Avenue, Cambridge CB3 0HE, UK}
\affiliation{INAF -- Osservatorio Astronomico di Brera, via Brera 28, I-20121 Milano, Italy}

\author[0000-0002-4622-6617] {Fengwu Sun}
\affiliation{Center for Astrophysics $|$ Harvard \& Smithsonian, 60 Garden St., Cambridge MA 02138 USA}

\author[0000-0003-4337-6211] {Jakob M.\ Helton}
\affiliation{Steward Observatory, University of Arizona, 933 N. Cherry Ave, Tucson, AZ 85721, USA}

\author[0000-0002-5500-4602] {Brittany E. Miles}
\altaffiliation{51 Pegasi B Fellow}
\affiliation{Steward Observatory, University of Arizona, 933 N. Cherry Ave, Tucson, AZ 85721, USA}

\author[0000-0002-5251-2943] {Mark S. Marley}
\affiliation{Department of Planetary Sciences, Lunar \& Planetary Laboratory, University of Arizona, Tucson, AZ 85721, USA}

\author[0000-0003-1487-6452] {Ben W. P. Lew}
\affiliation{Bay Area Environmental Research Institute, Moffett Field, CA 94035, USA}
\affiliation{NASA Ames Research Center, Moffett Field, CA 94035, USA}

\author[0000-0002-0834-6140] {Jarron M.\ Leisenring}
\affiliation{Steward Observatory, University of Arizona, 933 N. Cherry Ave, Tucson, AZ 85721, USA}

\author[0000-0002-8651-9879] {Andrew J.\ Bunker}
\affiliation{Department of Physics, University of Oxford, Denys Wilkinson Building, Keble Road, Oxford OX1 3RH, UK}

\author[0000-0002-1617-8917]{Phillip A.\ Cargile}
\affiliation{Center for Astrophysics $|$ Harvard \& Smithsonian, 60 Garden St., Cambridge MA 02138 USA}

\author[0000-0002-6719-380X] {Stefano Carniani}
\affiliation{Scuola Normale Superiore, Piazza dei Cavalieri 7, I-56126 Pisa, Italy}

\author[0000-0002-2929-3121] {Daniel J.\ Eisenstein}
\affiliation{Center for Astrophysics $|$ Harvard \& Smithsonian, 60 Garden St., Cambridge MA 02138 USA}

\author[0009-0003-7423-8660]{Ignas Juod\v{z}balis}
\affiliation{Kavli Institute for Cosmology, University of Cambridge, Madingley Road, Cambridge CB3 0HA, UK}
\affiliation{Cavendish Laboratory, University of Cambridge, 19 JJ Thomson Avenue, Cambridge CB3 0HE, UK}

\author[0000-0002-9280-7594] {Benjamin D.\ Johnson}
\affiliation{Center for Astrophysics $|$ Harvard \& Smithsonian, 60 Garden St., Cambridge MA 02138 USA}

\author[0000-0002-4271-0364] {Brant Robertson}
\affiliation{Department of Astronomy and Astrophysics, University of California, Santa Cruz, 1156 High Street, Santa Cruz CA 96054, USA}

\author[0000-0002-8224-4505] {Sandro Tacchella}
\affiliation{Kavli Institute for Cosmology, University of Cambridge, Madingley Road, Cambridge CB3 0HA, UK}
\affiliation{Cavendish Laboratory, University of Cambridge, 19 JJ Thomson Avenue, Cambridge CB3 0HE, UK}

\author[0000-0003-2919-7495] {Christina C. Williams}
\affiliation{NSF’s National Optical-Infrared Astronomy Research Laboratory, 950 North Cherry Avenue, Tucson, AZ 85719, USA}

\author[0000-0001-9262-9997] {Christopher N.\ A.\ Willmer}
\affiliation{Steward Observatory, University of Arizona, 933 N. Cherry Ave, Tucson, AZ 85721, USA}

\begin{abstract}

Large area observations of extragalactic deep fields with the James Webb Space Telescope (JWST) have provided a wealth of candidate low-mass L- and T-class brown dwarfs. The existence of these sources, which are at derived distances of hundreds of parsecs to several kiloparsecs from the Sun, has strong implications for the low-mass end of the stellar initial mass function, and the link between stars and planets at low metallicities. In this letter, we present a JWST/NIRSpec PRISM spectrum of brown dwarf JADES-GS-BD-9, confirming its photometric selection from observations taken as part of the JWST Advanced Deep Extragalactic Survey (JADES) program. Fits to this spectrum indicate that the brown dwarf has an effective temperature of 800-900K (T5 - T6) at a distance of $1.8 - 2.3$kpc from the Sun, with evidence of the source being at low metallicity ([M/H] $\leq -0.5$). Finally, because of the cadence of JADES NIRCam observations of this source, we additionally uncover a proper motion between the 2022 and 2023 centroids, and we measure a proper motion of $20 \pm 4$ mas yr$^{-1}$ (a transverse velocity of 214 km s$^{-1}$ at 2.25 kpc). At this predicted metallicity, distance, and transverse velocity, it is likely that this source belongs either to the edge of the Milky Way thick disk or the galactic halo. This spectral confirmation demonstrates the efficacy of photometric selection of these important sources across deep extragalactic JWST imaging. 
\end{abstract}

\keywords{Brown dwarfs(185)	 --- T dwarfs(1679) --- Halo stars(699) --- Infrared astronomy(786) --- James Webb Space Telescope(2291}

\section{Introduction} \label{sec:intro}

The last two years have seen significant progress in the understanding of cold brown dwarfs, astronomical objects that link stellar and exoplanet studies. Brown dwarfs have low masses (M$_{\ast}$ $< 0.07$ M$_{\odot}$, under the stellar hydrogen core fusion limit) and low effective temperatures ($T_{\mathrm{eff}} < 2500$ K), and they provide insight into star and planet formation, atmospheric physics, and the low-mass end of the stellar initial mass function. At such low temperatures and small sizes, brown dwarfs with late M, L, T, and Y spectral types are very faint, and so understanding these sources has historically been restricted to the solar neighborhood \citep{nakajima1995, kirkpatrick2011, mace2013, meisner2021}. Local late-T and Y dwarfs have masses of only $0.004 - 0.05$ M$_{\odot}$ (5 - 50 M$_{\mathrm{Jupiter}}$) and have metallicities that range between -0.5 to +0.3 dex \citep{line2017, leggett2021} ([M/H], where solar metallicity is given as [M/H] $= 0$), in agreement with expectations for their position within the Milky Way thin disk \citep{kilic2019, hallakoun2021}.

The discovery of brown dwarfs at larger distances, especially those in the Milky Way thick disk and even the halo, is crucial for our understanding of the history of star formation in the galaxy, and over the last decade several studies have uncovered members of this population, often using wide-area infrared observations from space \citep{burgasser2003b, loediu2010, murray2011, burningham2014, pinfield2014, zhang2019, schneider2020}. The James Webb Space Telescope (JWST), which has sensitive instruments that target the near- and mid-infrared, has been an ideal observatory to understand brown dwarfs both locally and even at hundreds to thousands of parsecs. In \citet{nonino2023}, \citet{wang2023}, \citet{hainline2024b}, and \citet{langeroodi2023}, photometric selection methods are introduced using JWST/NIRCam filters to separate brown dwarfs from galaxies in deep extragalactic surveys. \citet{hainline2024b} used exceptionally deep NIRCam imaging from the JWST Advanced Deep Extragalactic Survey \citep[JADES;][]{eisenstein2023, jades-doi} and the Cosmic Evolution Early Release Science (CEERS) surveys \citep{finkelstein2022} to find a sample of 21 T and Y (T$_{\mathrm{eff}}$ = 500K - 1200K) brown dwarf candidates at predicted distances between $0.1 - 4.2$ kpc. A subsample of seven of these candidates had observed proper motions (measured between HST, Spitzer, and JWST images of the sources taken across multiple years) which implied transverse velocities between 30 - 120 km s$^{-1}$, indicating that these objects were likely not extragalactic. 

These new samples of distant cold brown dwarf candidates identified by JWST are exciting, but require spectroscopy both to confirm their nature and understand their detailed atmospheric properties. In \citet{burgasser2024} and \citet{langeroodi2023}, the authors independently studied JWST/NIRSpec PRISM spectroscopy for a sample of three brown dwarfs observed as part of the Ultradeep NIRSpec and NIRCam ObserVations before the Epoch of Reionization \citep[UNCOVER;][]{bezanson2022} and Grism Lens-Amplified Survey from Space \citep[GLASS; ][]{treu2022} surveys. From the fits done by \citet{burgasser2024}, these brown dwarfs have distances between $0.87 - 4.5$ kpc, effective temperatures between $550 - 1110$K, and intriguingly, two of the sources, UNCOVER-BD-1 and UNCOVER-BD-3, are best-fit with low-metallicity models ([M/H] $\leq -0.5$). The authors conclude that there is high probability that they reside in the Milky Way thick disk or halo. Thick disk or halo brown dwarfs are expected to have low metallicities and high transverse velocities, but they are very rare \citep{faherty2009, kilic2019, hallakoun2021, meisner2023}. Given the importance in how these objects probe the early history of star formation in the Milky Way, and the lack of constraints on models of low-metallicity brown dwarf atmospheres, it is imperative to increase the number of these sources with observed spectroscopy. 

In this paper, we describe updated photometric and spectroscopic observations of JADES-GS-BD-9, a brown dwarf first identified in \citet{hainline2024b}. We present a JWST/NIRSpec PRISM spectrum for the source, which we fit with both standard brown dwarf templates and multiple brown dwarf atmospheric models. In addition, because of the cadence of the JADES NIRCam photometric observations, we identify a proper motion for this source, and estimate a transverse velocity to explore the position of this brown dwarf within the Milky Way. We describe the data we use for this study in Section \ref{sec:observations}, we discuss our fits to this data in Section \ref{sec:analysis}, including the spectroscopic (\S \ref{subsec:nirspec}) and proper motion (\S \ref{subsec:nircam}) fits, we discuss the implications of these fits in Section \ref{sec:results}, and we conclude in Section \ref{sec:conclusion}.

\section{Observations and Data Reduction} \label{sec:observations}

\begin{figure*}
  \centering
  \includegraphics[width=0.99\linewidth]{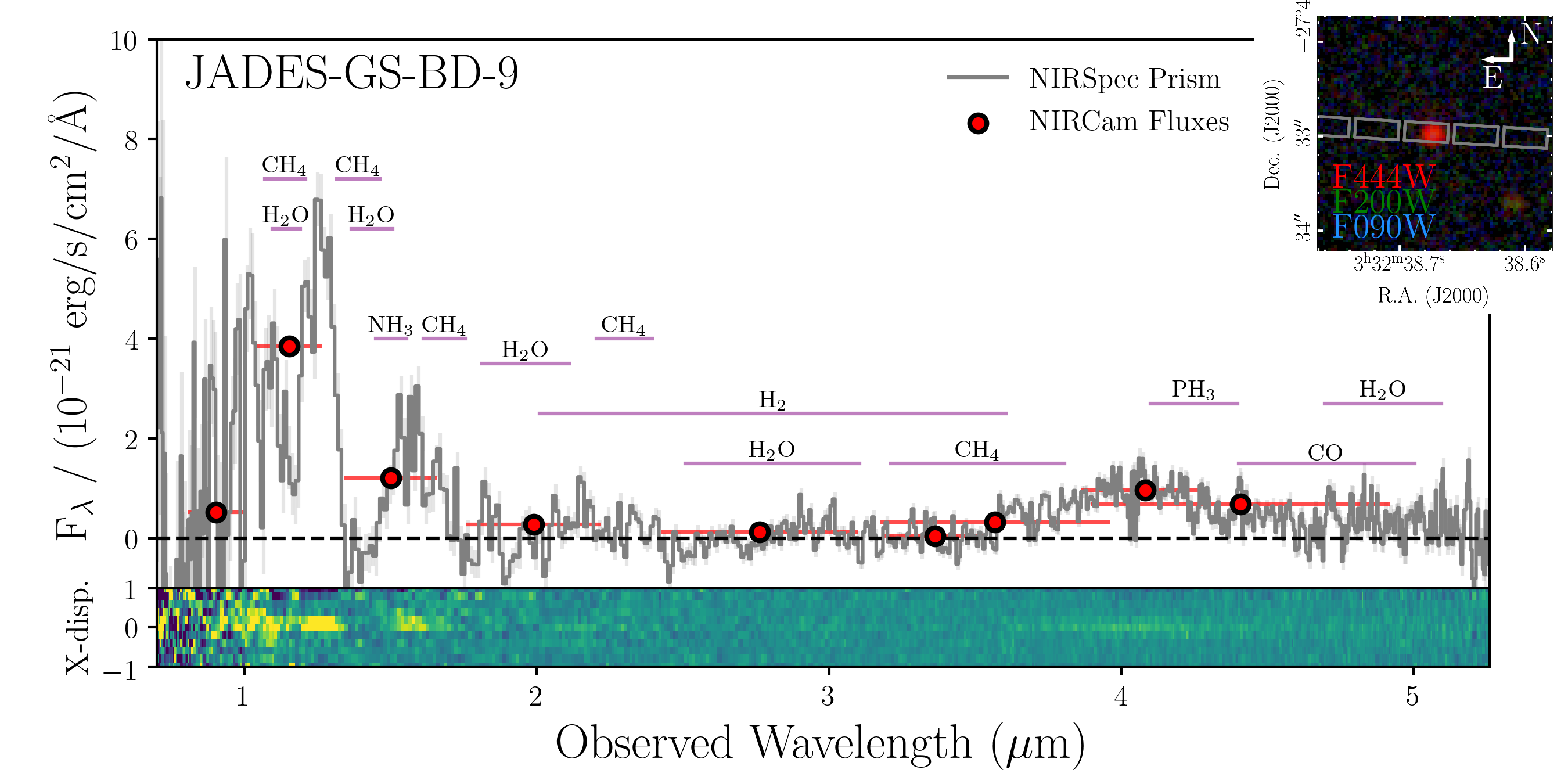}
  \caption{(Top) 3 pixel boxcar extracted 1D NIRSpec PRISM spectrum in grey, with the NIRCam fluxes from \citet{hainline2024b} plotted as red points. We plot the strong molecular absorption bands that shape the observed spectrum, labeled by their species, with purple horizontal lines. In an inset we show the $2^{\prime\prime} \times 2^{\prime\prime}$ RGB thumbnail for the brown dwarf, with the NIRSpec MSA slit positions overplotted. (Bottom) 2D NIRSpec PRISM spectrum for JADES-GS-BD-9.}
  \label{fig:2d-1d-spectrum}
\end{figure*}

JADES-GS-BD-9 (J2000 R.A.: $53.161140$, DEC: $-27.809163$) was initially selected from JADES photometry using the color and magnitude selection criteria as described in \citet{hainline2024b}. While this source appears just south of the Hubble Ultra Deep Field, and has extensive deep HST/ACS and WFC3 imaging, it is not detected in mosaics from either instrument. In \citet{hainline2024b}, the authors fit the NIRCam model photometry for the source to estimate the distance (1430 - 2080 pc, depending on the models used), effective temperature (800 - 900K), and other atmospheric properties. The source was further spectroscopically observed with the JWST/NIRSpec multi-shutter array (MSA) \citep{jakobsen2022,ferruit2022} as part of the JADES Medium Survey (Program ID 1180, PI Eisenstein) in October of 2023 (Obs136). These observations were assigned as compensation for previously failed observations \citep[due to electric short circuits;][]{rawle2022}. Therefore, JADES used new JWST/NIRCam data to re-design the original, failed mask (Obs30).

The observations use four filter/disperser combinations: low-resolution clear/PRISM with $R=30\text{--}300$ and medium-resolution f070lp/g140m, f170lp/g235m, f290lp/g395m, with $R=700\text{--}1500$ \citep{jakobsen2022}. Both the PRISM and the ensemble of grating observations cover the full wavelength range of NIRSpec, from 0.7 to 5.2~$\mu$m. The effective resolutions are higher than the nominal values, because at blue wavelengths the source is convolved with the point-spread-function (the PSF is significantly narrower than the width of the NIRSpec/MSA micro-shutters (0.2$^{\prime\prime}$). In this work, we use only the PRISM data, which provides a larger signal-to-noise ratio per spectral resolution element, as we do not measure any significant flux in the grating spectra for the source. The PRISM data were obtained using two dithers and a three-nod pattern for each dither, giving a total of six exposures. Each exposure consisted of a single integration of 17 groups, in NRSIRS2 readout mode \citep{rauscher+2012,rauscher+2017}, adding up to a total exposure time of 7528 sec ($\sim2.1$ hours). These data were reduced using the pipeline developed by the ESA NIRSpec Science Operations Team (SOT) and Guaranteed Time Observations (GTO) NIRSpec teams, as described in  \citet{bunker2023} and \citet{deugenio2024}. Here we remark that we used nodding for subtracting the background, and path-loss flux corrections appropriate for a point source.

We plot the observed 2D and 1D PRISM spectra at  1 - 5$\mu$m for JADES-GS-BD-9 in Figure \ref{fig:2d-1d-spectrum}. In the 2D spectrum, we can see the peaks at $1.0\mu$m, $1.4\mu$m, $2.2\mu$m commonly associated with brown dwarf near-IR spectra, which arise due to absorption in the brown dwarf atmospheres from H$_2$O and CH$_4$ \citep{marley1996, burrows1997, allard1997}. We also see the faint trace of a broad red feature at $\sim4\mu$m. In the 1D plot, we overplot the JADES NIRCam photometry from \citet{hainline2024b} in red and find that these agree with the observed spectrum, validating the path-loss corrections and flux calibration. The observed spectrum confirms that JADES-GS-BD-9 is a brown dwarf. The shape of the spectrum is very similar to the NIRCam-selected brown dwarfs presented in \citet{burgasser2024}, in particular UNCOVER-BD-1 and UNCOVER-BD-2. In an inset above the spectrum we show the $2^{\prime\prime} \times 2^{\prime\prime}$ RGB thumbnail for the source, composed of images taken with the F090W (blue), F200W (green), and F444W (red) filters. In this inset, we overplot the NIRSpec MSA slit position used for targeting this source. As measured and discussed in \citet{hainline2024b}, the source is unresolved. 

We also explore the NIRCam position of this source as measured in the 2022 and 2023 JADES observations. The NIRCam observations that were used to identify JADES-GS-BD-9 were taken in late September 2022, and additional JADES observations of the source at the end of September 2023. These data, including their observational setup and reduction are described in \citet{rieke2023} and \citet{eisenstein2023}. For this study, we primarily focus on the NIRCam images in the F115W (with signal-to-noise ratio (SNR) of 19.89 in \citealp{hainline2024b}) and F444W filters (SNR = 18.82), as these are the filters with the largest SNR. 

\section{Data Analysis} \label{sec:analysis}

We fit the observed NIRSpec spectrum of JADES-GS-BD-9 with both observed brown dwarf spectral standards as well as multiple atmospheric models. In addition, we explored the proper motion of the source between the 2022 and 2023 NIRCam observations. 

\subsection{Spectroscopic Fits} \label{subsec:nirspec}

\begin{deluxetable}{l c}
\tabletypesize{\footnotesize}
\tablecolumns{3}
\tablewidth{0pt}
\tablecaption{Observational Properties and Best-Fit Parameters\label{tab:spectrum-fit-parameters}}
\tablehead{
\colhead{Parameter} &  \colhead{JADES-GS-BD-9}}
\startdata
  R.A. (degrees) & 53.161140 \\ 
  DEC (degrees) & -27.809163 \\
  t$_{\mathrm{exp,NIRSpec}}$ & 7528 s\\
  Spec. Type & T5-T6\\
\hline
  \multicolumn{2}{c}{{\tt Sonora Elf Owl}} \\
\hline
    $\chi^2_{\mathrm{red}}$ & 4.56 \\
    $T_{\mathrm{eff}}$ & $900$K  \\
    $\log{(g)}$ & 5.0 \\
    {[}M/H{]} & -0.5  \\
    distance & 2255 pc  \\
    radius & 0.093 R$_{\odot}$ \\
\hline
\multicolumn{2}{c}{{\tt ATMO2020++}} \\
\hline
    $\chi^2_{\mathrm{red}}$ & 4.77 \\
    $T_{\mathrm{eff}}$ & $800$K \\
    $\log{(g)}$ & 3.5  \\
    {[}M/H{]} & -1.0  \\
    distance & 1823 pc  \\
    radius & 0.079 R$_{\odot}$ \\
\hline
\multicolumn{2}{c}{\tt LOWZ} \\
\hline
    $\chi^2_{\mathrm{red}}$ & 4.51 \\
    $T_{\mathrm{eff}}$ & $800$K \\
    $\log{(g)}$ & 5.25  \\
    {[}M/H{]} & -0.5  \\
    distance & 2056 pc  \\
    radius$^{a}$ & 0.103 R$_{\odot}$ \\
\enddata
\tablecomments{$^{a}$We assumed 1 Jupiter radius for the LOWZ models.}
\end{deluxetable}

\begin{figure*}
  \centering
  \includegraphics[width=0.48\linewidth]{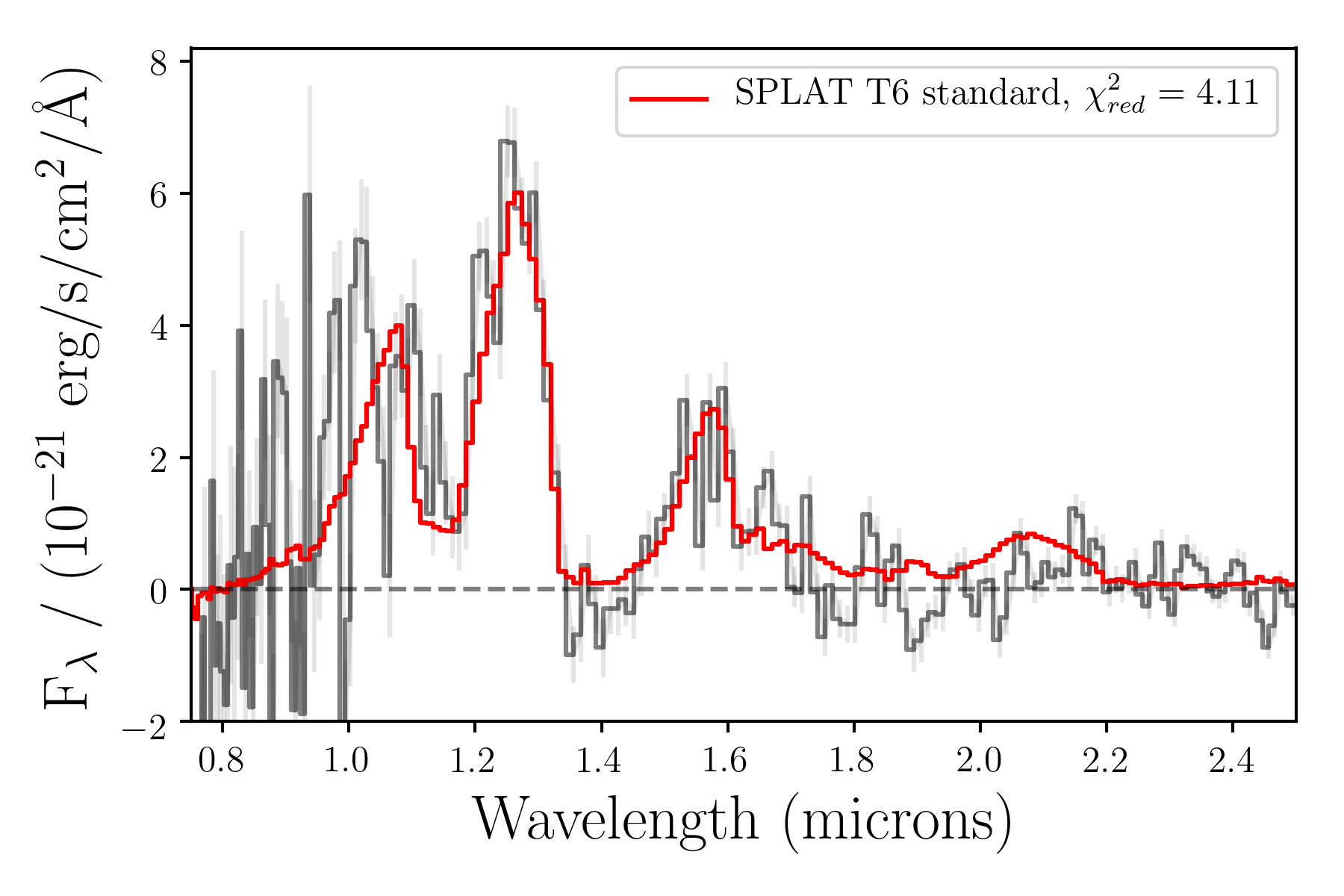}
  \includegraphics[width=0.48\linewidth]{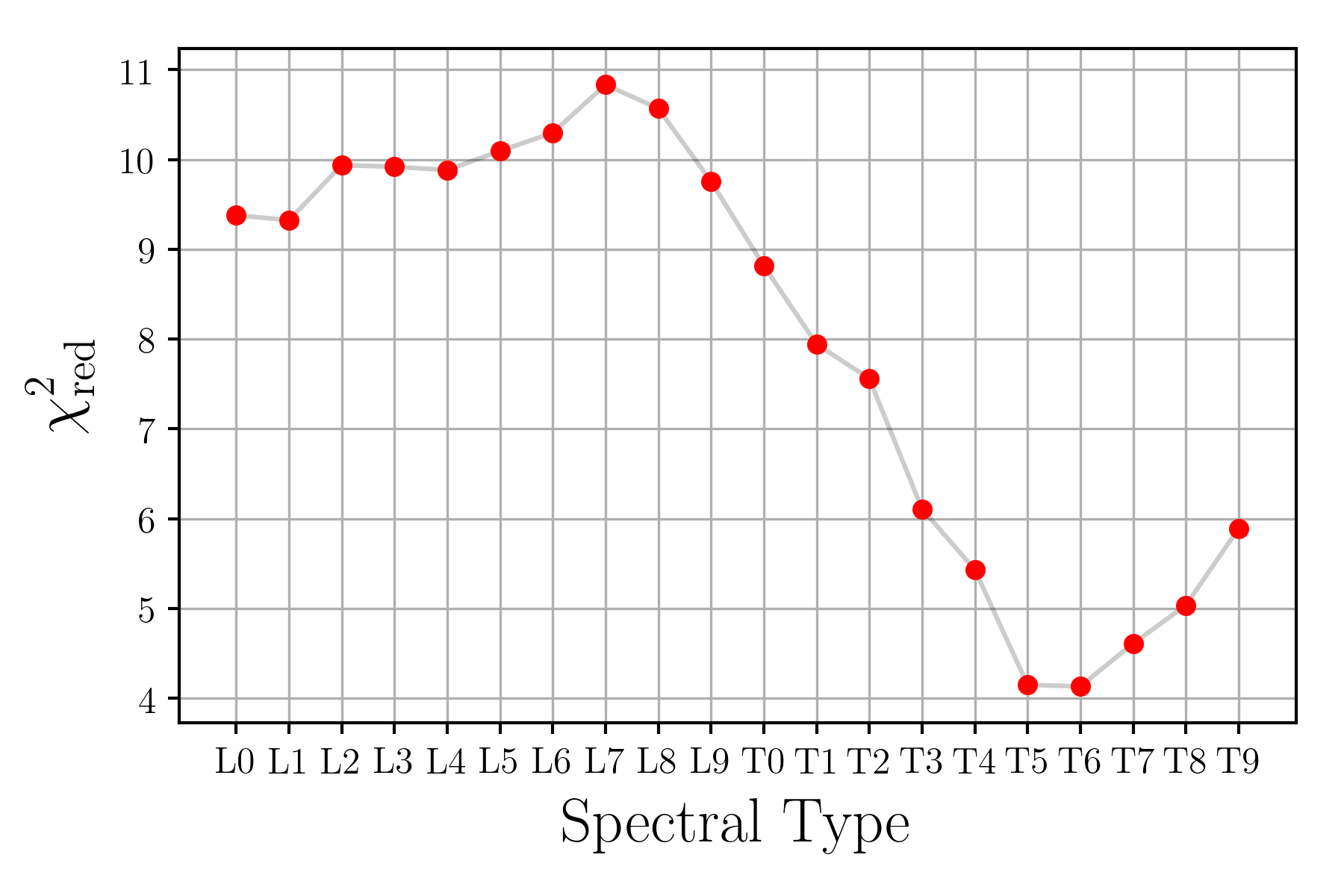}
  \caption{(Left) 1D NIRSpec/PRISM spectrum compared to the best-fit standard brown dwarf from SPLAT. (Right) Reduced $\chi^2$ compared to spectral type for the L and T standards from SPLAT.}
  \label{fig:1d-spectrum-splat}
\end{figure*}

To understand the spectral type for JADES-GS-BD-9, we compared the NIRSpec spectrum for this source to observed brown dwarf standards using the SpeX Prism Library Analysis Toolkit \citep[SPLAT;][]{burgasser2017}. The SPLAT spectra extend to $\lambda < 2.5\mu$m, and we measured the reduced $\chi^2$ between the observed NIRSpec spectrum at these wavelengths and the L and T dwarf standard spectra. We plot the fit with the lowest reduced $\chi^2$, a T6 brown dwarf, in the left panel of Figure \ref{fig:1d-spectrum-splat}. The T6 standard agrees quite well with the JADES-GS-BD-9 spectrum, although the relatively faint peak at 2.0 - 2.2$\mu$m in the standard spectrum does not appear in the observed spectrum. In addition, the JADES-GS-BD-9 spectrum also resembles the near-IR spectrum for the T5 brown dwarf ULAS J124639.32+032314.2 (UL1246) presented in \citet{zhang2019}, as well as the T5 brown dwarf 2MASS J15031961+2525196 \citep[2M1503;][]{burgasser2003c,burgasser2006}, although both of those sources have significantly stronger flux observed at 2.0 - 2.2$\mu$m than we see in JADES-GS-BD-9. In the right panel of the Figure \ref{fig:1d-spectrum-splat} we show the reduced $\chi^2$ values for all of the SPLAT comparisons as a function of spectral type, demonstrating that while the fit to the T6 standard returns the overall minimum reduced $\chi^2$, the T5 and T6 spectra had similar best-fit minimum values. This agrees with the type range presented in \citet{hainline2024b}, T6–T7, from model fits to the photometry. 

We also fit the JADES-GS-BD-9 spectrum to a suite of brown dwarf model packages as a way of understanding the distance, effective temperature, and metallicity of the source. We used three model packages: ATMO2020++ \citep{phillips2020, meisner2023}, LOWZ \citep{meisner2021}, and SONORA Elf Owl \citep{mukherjee2024}. Each of these models was generated from a 1D radiative-convective equilibrium code, and are cloud-free. The ATMO2020++ models have an effective temperature range of $250 - 1200$K ($\Delta T_{\mathrm{eff}} = 25$K between 250K and 300K, $\Delta T_{\mathrm{eff}} = 50$K between 300K and 500K, and then $\Delta T_{\mathrm{eff}} = 100$K between 500K and 1200K), a specific gravity range of log(g) $= 2.5 - 5.5$, ($\Delta \mathrm{log(g)} = 0.5$) and allow for metallicity values of [M/H] $ = -1, -0.5, 0.0$. The LOWZ models span $500 - 1600$K ($\Delta T_{\mathrm{eff}} = 50$K between 500K and 1000K, and then $\Delta T_{\mathrm{eff}} = 100$K between 1000K and 1600K), log(g) $= 3.5 - 5.25$ ($\Delta \mathrm{log(g)} = 0.5$ between 3.5 and 5.0, and $\Delta \mathrm{log(g)} = 0.25$ between 5.0 and 5.25), and, have a metallicity range between between -$2.5$ and $+1.0$, with $\Delta$[M/H] = 0.5. Finally, the Sonora Elf Owl models have $T_{\mathrm{eff}} = 500 - 1300$K ($\Delta T_{\mathrm{eff}} = 25$K between 575 and 600K, and then $\Delta T_{\mathrm{eff}} = 50$K between 600K and 1000K), log(g) $= 3.0 - 5.5$ ($\Delta \mathrm{log(g)} = 0.25$), and also allow for metallicity values of [M/H] $ = -1, -0.5, 0.0, +0.5, +0.7$, and $+1.0$. For each code, we match the resolution of the model brown dwarf spectra to the NIRSpec PRISM resolution and fit to the observed spectrum for JADES-GS-BD-9, and we report the fits with the minimum $\chi^2$. 

To estimate the distances from the models, we use the best-fit normalization of the model to the observed spectrum which scales with $(R/D)^2$, where $R$ is the radius of JADES-GS-BD-9, and $D$ is the distance to the object. For the ATMO2020++ distance estimation, we use the solar metallicity evolution models from ATMO2020 \citep{phillips2020}, where the predicted radii for each model at a range of ages, and we adopt the radius given for an age of $\log{\mathrm{(age/yr)}}$ = 10. We also looked at the results we would get if we adopted an age for JADES-GS-BD-9 of $\log{\mathrm{(age/yr)}}$ = 7.9, where the radius and the distance are 2\% larger. For the Sonora Elf Owl models, we use the radius at the best-fit effective temperature and log(g) provided by the Sonora Bobcat evolutionary models \citep{marley2021}, as described in \citet{hainline2024b}, and adopted an age of $\log{\mathrm{(age/yr)}}$ = 9.2. We note that the Sonora Bobcat models are calculated at solar metallicity, and going to lower values would increase the assumed radius (and decrease the estimated distance) by only a few percent. For both the ATMO2020++ and Sonora Elf Owl models, we adopt an age expected for thick disk or halo brown dwarfs and stars. For the LOWZ models, we fix the radius of JADES-GS-BD-9 to one Jupiter radius. This value agrees with what we find from our fits using the other models: the best-fit radius of the brown dwarf predicted from the Sonora Bobcat Models is $\sim 10$\% larger, and the radius predicted from the LOWZ models is only $\sim 6$\% larger.

\begin{figure*}
  \centering
  \includegraphics[width=0.99\linewidth]{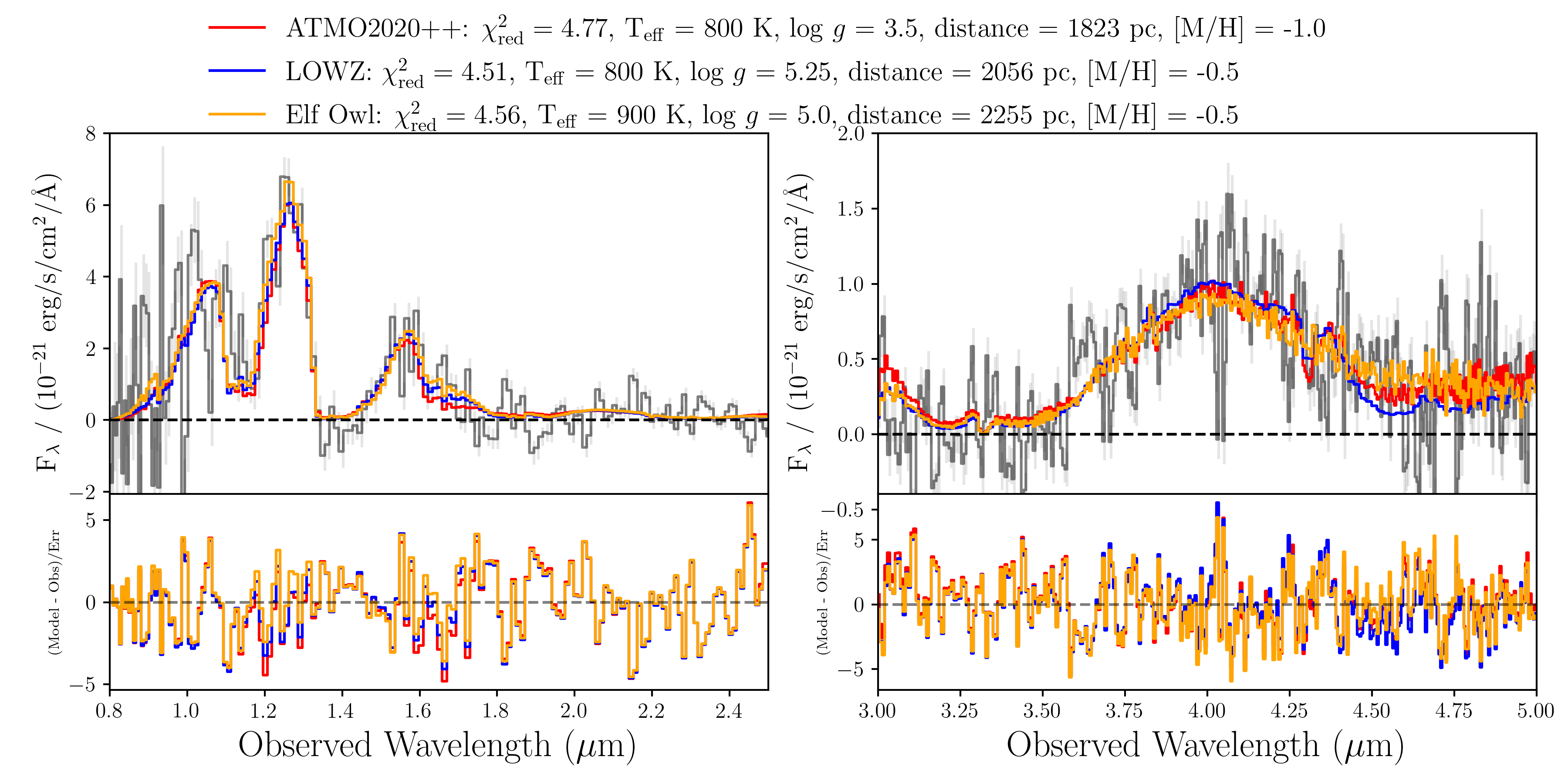}
  \caption{(Top) 1D NIRSpec PRISM spectrum for JADES-GS-BD-9, with ATMO2020++ (red) LOWZ (blue), and Sonora Elf Owl (orange) fits. The best-fit parameters are shown in the legend above. We split the spectrum to help focus on the near-IR (left) and 3 - 5$\mu$m region (right). (Bottom) Residuals for the three fits. All three models match the observed spectrum within the uncertainties, and all three predict a subsolar metallicity for the brown dwarf.}
  \label{fig:1d-spectrum-ATMO-LOWZ}
\end{figure*}

We plot the best-fit model predictions along with the NIRSpec spectrum for JADES-GS-BD-9 in Figure \ref{fig:1d-spectrum-ATMO-LOWZ}, along with the residuals below. In the legend, and in Table \ref{tab:spectrum-fit-parameters}, we provide the brown dwarf parameters derived from the multiple model fits. 

\subsection{NIRCam-Derived Proper Motions} \label{subsec:nircam}

\begin{figure*}
  \centering
  \includegraphics[width=0.8\linewidth]{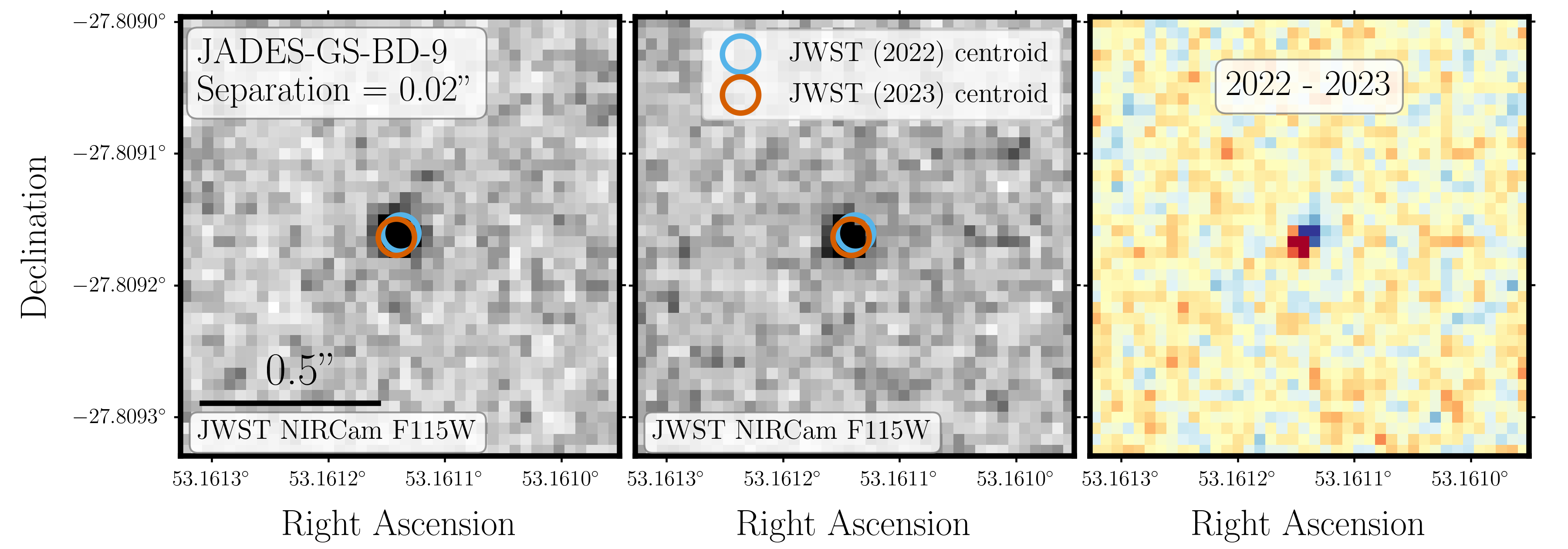}\\
  \vspace{0.5cm}
  \includegraphics[width=0.8\linewidth]{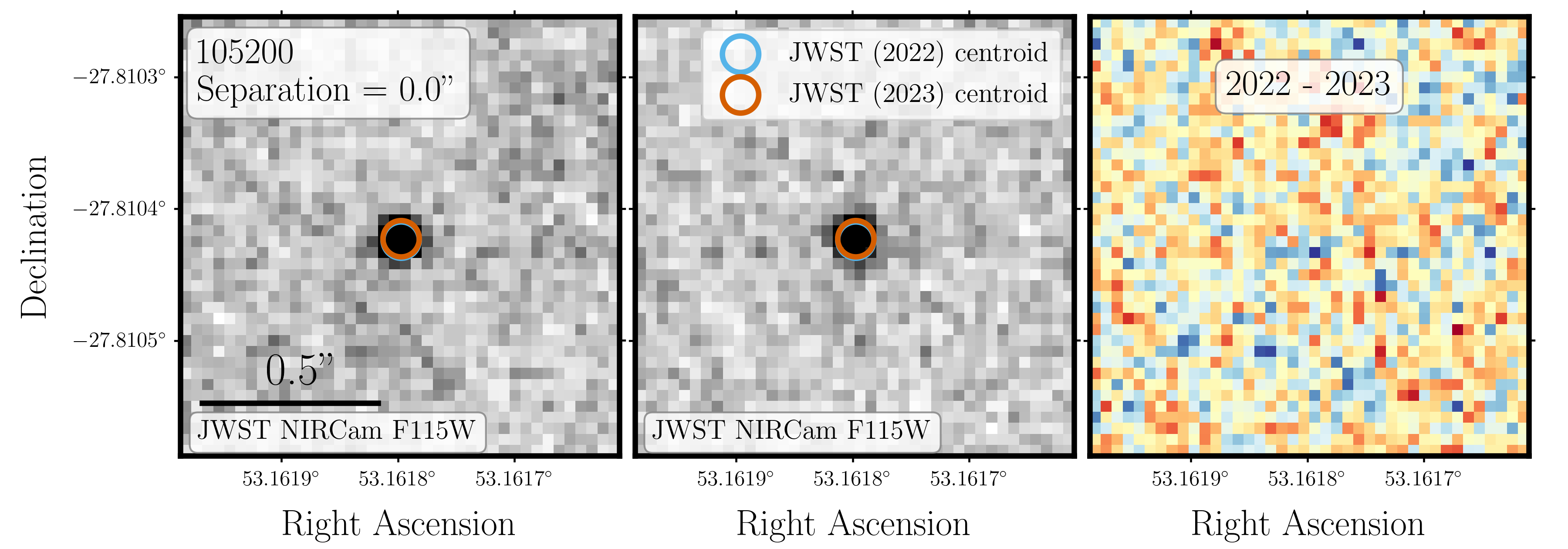}
  \includegraphics[width=0.8\linewidth]{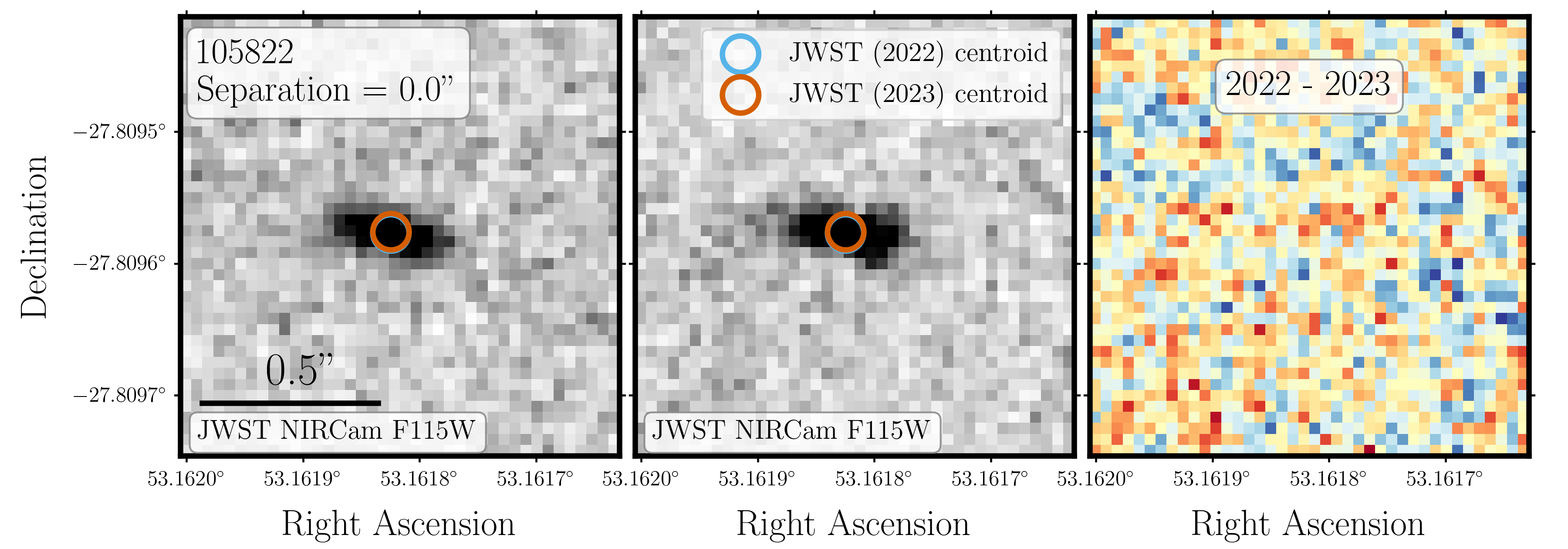}
  \includegraphics[width=0.8\linewidth]{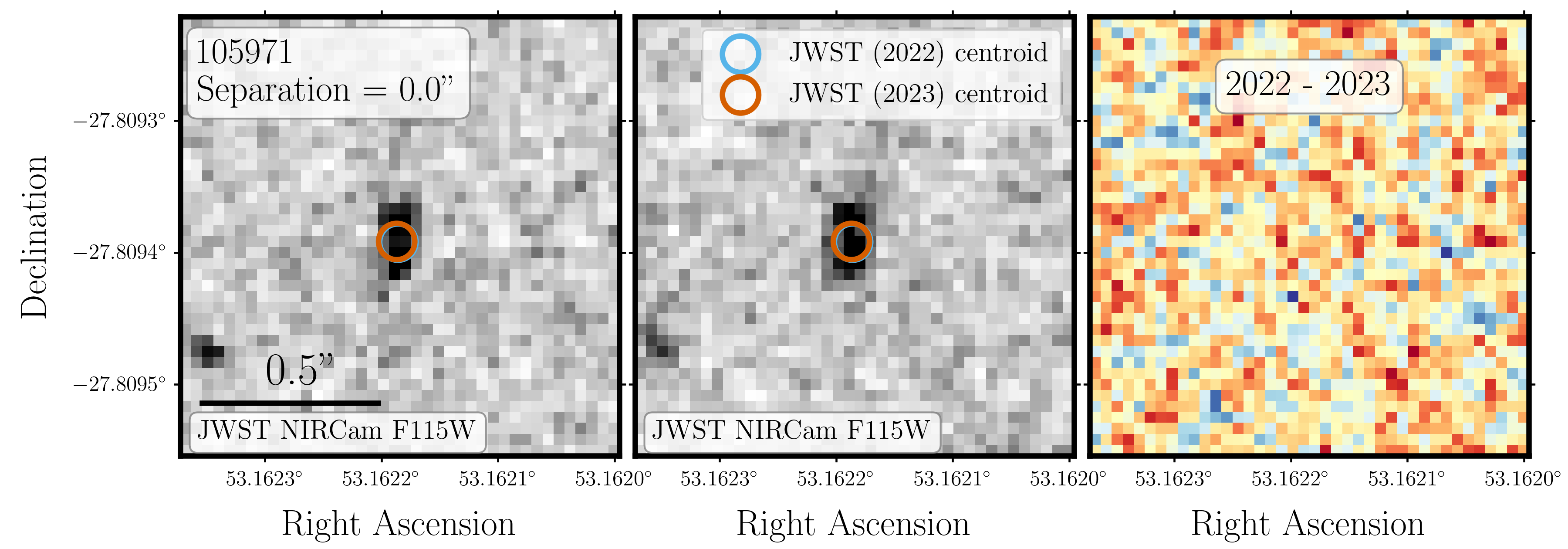}
  \caption{A comparison of the position of JADES-GS-BD-9 (first row) and three nearby bright galaxies (second, third, and fourth rows) in the 2022 JADES images (left column), 2023 JADES images (middle column), and in a difference image of the two years (right column). We can see that while JADES-GS-BD-9 has a proper motion shown by the changing centroid position and positive and negative flux in the difference image, the other nearby galaxies show no motions or significant residuals.}
  \label{fig:proper-motion}
\end{figure*}

The NIRCam observations that were used to initially identify JADES-GS-BD-9 were taken in late September 2022, and additional JADES observations were taken of the JADES Deep region where this source resides at the end of September 2023. These observations allow us to measure the proper motion for this source, which can be used to estimate a transverse velocity and better understand where JADES-GS-BD-9 resides within the Milky Way. In Figure \ref{fig:proper-motion} we plot the 2022 (left column) and 2023 (middle column) images in F115W for JADES-GS-BD-9 (top row). A positional offset for the source can be seen by comparing the 2022 and 2023 observations, which we estimate by fitting the centroid of the brown dwarf in each image. From the F115W images, we calculate a separation of 0.016$^{\prime\prime}$$\pm0.002^{\prime\prime}$ between the 2022 and 2023 positions. We note that the uncertainty of separation is estimated as the PSF FWHM divided by the source SNR. We additionally calculate the offsets measured from the F444W images, which have a larger PSF, and find a separation of 0.023$^{\prime\prime}$$\pm0.008^{\prime\prime}$. We adopt the mean of these measurements and present a proper motion of $20 \pm 4$ mas yr$^{-1}$ to the southeast for the source (a proper motion of $11 \pm 2$ mas yr$^{-1}$ to the east, and $16 \pm 3$ mas yr$^{-1}$ to the south). In Figure \ref{fig:proper-motion}, we plot the 2022 NIRCam F115W thumbnail for the source in the top left panel, the 2023 thumbnail in the top middle panel, and the 2023 thumbnail subtracted from the 2022 thumbnail in the top right panel. The color gradient seen in the difference image from the northwest (positive flux) to southeast (negative flux) demonstrates the proper motion of JADES-GS-BD-9. 

To assess the significance of this result, we explored the on-sky separation between the 2022 and 2023 centroid positions for a sample of 8 bright (defined as having $>10\sigma$ detections in at least three NIRCam photometric bands) sources within 10$^{\prime\prime}$ of JADES-GS-BD-9. The average RA offset we measure for these sources in the F115W image is only $0.8$ mas, and the average DEC offset we measure is only $-0.7$ mas. Combining these in quadrature, we measure an average offset of $1.0$ mas, and our measurement for JADES-GS-BD-9 is significantly in excess of this value. In addition, these sources do not show significant residuals in the difference images, unlike what we see for JADES-GS-BD-9. We plot three example nearby galaxies to JADES-GS-BD-9 in the bottom three rows of Figure \ref{fig:proper-motion}, and find no separation in the 2022 and 2023 positions, and no significant residual in the difference images.

From this proper motion we can calculate the transverse velocity $v_T = 4.74\mu d$, where $\mu$ is the proper motion in arcseconds per year, and $d$ is the distance to the source in pc. Adopting the Sonora Elf Owl distance (2255 pc), the transverse velocity of this source is $v_T \sim 214$km s$^{-1}$. If we instead adopt the smaller distance from the ATMO2020++ models, we estimate that $v_T \sim 172$km s$^{-1}$.

\section{Results} \label{sec:results}

From the observations presented here, we can confirm that JADES-GS-BD-9 is a brown dwarf. Excitingly, the spectral fits and proper motion for JADES-GS-BD-9 provide evidence for the brown dwarf being a member of the Milky Way thick disk or halo. As seen in Figure \ref{fig:1d-spectrum-splat}, while a T6 standard replicates the features seen between $1 - 1.6\mu$m in the NIRspec spectrum, there is less agreement between $2.0 - 2.3 \mu$m. Figure \ref{fig:1d-spectrum-ATMO-LOWZ} demonstrates that the best-fit subsolar ATMO2020+, LOWZ and Sonora Elf Owl models have significantly fainter predicted flux in this wavelength range which better matches the observed spectrum. At lower metallicities, the pressure-induced opacity of molecular hydrogen serves to reduce the flux at this wavelength \citep{saumon2012, marley2021}. This effect is similar to what is seen in the NIRSpec spectrum for UNCOVER-BD-1 from \citet{burgasser2024}, although that source is best fit with higher temperature models ($T_{\mathrm{eff}} = 1100 - 1500$K). 

The second piece of evidence for the position of JADES-GS-BD-9 in the Milky Way comes from the distance and observed proper motion. \citet{hainline2024b} present observations of proper motion for seven of the photometrically-selected candidate brown dwarfs, all at best-fit distances of $<800$ pc. The implied transverse velocities were all $<120$ km s$^{-1}$. The model fits to JADES-GS-BD-9 suggest the object is potentially located at twice this maximum distance and as a result has a significantly higher transverse velocity. Taken together, our estimates of the metallicity, distance, and transverse velocity for JADES-GS-BD-9 are in agreement with expectations for thick disk or halo brown dwarfs \citep{faherty2009, kilic2019, hallakoun2021, meisner2021}. 

For stars observed with \textit{Gaia}, broad kinematic cuts on total velocity ($v_{\mathrm{tot}}$) made to select thin-disk ($v_{\mathrm{tot}} < 50$ km s$^{-1}$), thick disk ($70 < v_{\mathrm{tot}} < 180$ km s$^{-1}$), and halo ($v_{\mathrm{tot}} > 200$ km s$^{-1}$) stars resulted in very different color-magnitude diagram distributions \citep[see Figures 21 and 22 in][]{gaia2018}. These cuts were chosen based on observations of the Milky Way from \citet{bensby2014}. The highest-velocity \textit{Gaia} subsample had both an extended horizontal branch and two distinct main sequences, and the authors model these with isochrones with [M/H] $= -1.3$, and a stellar age of 13 Gyr as well as one with [M/H] $= -0.5$, and a stellar age of 11 Gyr. As the transverse velocity we measure is a lower limit on the total velocity, we find that JADES-GS-BD-9 has observed kinematics and a best-fit metallicity in agreement with the latter population seen with \textit{Gaia}.

Intriguingly, we find absorption in the PRISM spectrum at 4.2 - 4.5$\mu$m for JADES-GS-BD-9, as plotted in the upper right panel of Figure \ref{fig:1d-spectrum-ATMO-LOWZ}. This narrow feature, which is also seen in the spectrum of UNCOVER-BD-3, was interpreted as evidence for phosphine (PH$_3$) absorption in \citet{burgasser2024}. These authors conclude that while CO$_2$ absorption can also affect this wavelength region, at the low temperature ($T_{\mathrm{eff}} = 600$K) they estimate for the source, phosphine absorption is likely the more significant contributor at these wavelengths. Phosphine absorption has been seen in spectra of both Jupiter and Saturn, but it has not been observed in most late-type brown dwarfs. In \citet{beiler2024}, the authors took an independent look at the NIRSpec spectrum for UNCOVER-BD-3 and other objects in their sample conclude that standard atmospheric models, such as the LOWZ models used in this study, overpredict the atmospheric phosphine abundance, and conclude that the observed feature is likely caused by CO$_2$. At the low SNR levels of both the UNCOVER-BD-3 and JADES-GS-BD-9 spectra, it is difficult to disentangle the effects of CO$_2$ and phosphine at these wavelengths. Further deep spectroscopic observations of these distant brown dwarfs are required to help constrain models and test the supposition in \citet{burgasser2024} that this phosphine absorption is perhaps common in sub-solar metallicity brown dwarfs. 

JADES-GS-BD-9 joins the three sources from \citet{burgasser2024} and \citet{langeroodi2023} as the four farthest brown dwarfs with spectroscopic confirmation. In all four cases, these sources were initially targeted for spectroscopy because they were thought to be red galaxies or active galactic nuclei based on their NIRCam photometry \citep{langeroodi2023, deugenio2024}. As was demonstrated in \citet{hainline2024b}, there are several planned or existing large area JWST surveys at depths that will select a significant number of distant brown dwarf candidates. While photometric fits can be used to estimate effective temperature, age, and distance, the analysis of the near-IR spectra for these sources is the only way to fully understanding their atmospheric properties and metallicity. 

\section{Conclusions} \label{sec:conclusion}

We present the NIRSpec PRISM spectrum for JADES-GS-BD-9, a low-temperature (T5-T6, $T_{\mathrm{eff}} = 800 - 900$K) brown dwarf originally selected photometrically from deep extragalactic observations. Fits to the spectrum indicate that this brown dwarf is at low metallicity ([M/H]$ \leq -0.5$), with a distance of $\sim2$kpc from the Sun. We additionally measure a proper motion of 20 mas yr$^{-1}$ between JWST/NIRCam observations of the source taken in 2022 and 2023. At the predicted distance, this implies a transverse velocity for JADES-GS-BD-9 of $172 - 214$ km s$^{-1}$. 

This spectrum helps to validate the photometric selection of low-temperature brown dwarfs from deep extragalactic survey observations, as outlined in \citet{nonino2023}, \citet{langeroodi2023} and \citet{hainline2024b}. The number densities of these sources are in agreement with brown dwarf and low-mass model predictions within the Milky Way \citep{hainline2024b}. As such, given the current and planned surveys over large areas with JWST/NIRCam coverage, it is likely that significantly more kpc-distance brown dwarfs will be found in the coming years. Given their potential position at the edges of our galaxy, spectroscopic follow-up is vital for better understanding these sources and constraining models for low-metallicity brown dwarf atmospheres. Distant, low-metallicity brown dwarfs like JADES-GS-BD-9 also shed light on low-mass star formation early in the history of the Milky Way. 

\bigskip
This work is funded through the JWST/NIRCam contract to the University of Arizona, NAS5-02015, and JWST Program 3215. The JWST data presented in this article were obtained from the Mikulski Archive for Space Telescopes (MAST) at the Space Telescope Science Institute. The specific observations analyzed can be accessed via doi:10.17909/8tdj-8n28. AJB acknowledges funding from the ``FirstGalaxies'' Advanced Grant from the European Research Council (ERC) under the European Union’s Horizon 2020 research and innovation programme (Grant agreement No. 789056). B.E.M. was supported by the Heising–Simons Foundation 51 Pegasi b Postdoctoral Fellowship. S.C acknowledges support by European Union’s HE ERC Starting Grant No. 101040227 - WINGS. FDE and IJ acknowledge support by the Science and Technology Facilities Council (STFC), by the ERC through Advanced Grant 695671 ``QUENCH'', and by the UKRI Frontier Research grant RISEandFALL. DJE is supported as a Simons Investigator and by JWST/NIRCam contract to the University of Arizona, NAS5-02015. ST acknowledges support by the Royal Society Research Grant G125142. The research of CCW is supported by NOIRLab, which is managed by the Association of Universities for Research in Astronomy (AURA) under a cooperative agreement with the National Science Foundation. BER acknowledges support from the NIRCam Science Team contract to the University of Arizona, NAS5-02015, and JWST Program 3215. The authors acknowledge use of the lux supercomputer at UC Santa Cruz, funded by NSF MRI grant AST 1828315.

\vspace{5mm}
\facilities{JWST(NIRCam and NIRSpec)}

\software{astropy \citep{astropy:2013, astropy:2018, astropy:2022}
          }

\bibliography{bdbib}{}
\bibliographystyle{aasjournal}

\end{document}